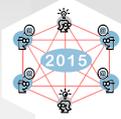

# Realistic, Extensible DNS and mDNS Models for INET/OMNeT++

Andreas Rain        Daniel Kaiser        Marcel Waldvogel

University of Konstanz, Konstanz, Germany

*<first>.<last>@uni-konstanz.de*

*Abstract*—**The domain name system (DNS) is one of the core services in today's network structures. In local and ad-hoc networks DNS is often enhanced or replaced by mDNS. As of yet, no simulation models for DNS and mDNS have been developed for INET/OMNeT++. We introduce DNS and mDNS simulation models for OMNeT++, which allow researchers to easily prototype and evaluate extensions for these protocols.**

**In addition, we present models for our own experimental extensions, namely *Stateless DNS* and *Privacy-Enhanced mDNS*, that are based on the aforementioned models. Using our models we were able to further improve the efficiency of our protocol extensions.**

## I. INTRODUCTION

The domain name system specified by RFC 1035 [1] is used to name and share resources, making it a crucial part of the world wide web as we know it. Although DNS in general is a well-researched topic, its extensible nature continuously provides grounds for new research. With a simulation model, behavior and performance of DNS and extensions to DNS can be evaluated more rapidly without the need of using real systems. As an example, it may be easier to extend the simulation model by new caching strategies and evaluate these strategies by predefined measurements and behavioral studies than to integrate the strategies into an existing system and test it using real clients. It is also easier to instrument the simulation to capture statistical information, than in a real world system. We use the proposed models, e.g. to evaluate our *Stateless DNS* technique [2], which utilizes caching behaviors of DNS servers to answer queries without itself holding state.

We also present a model for multicast DNS as specified in RFC 6762 [3], which is widely used in local networks for which a dedicated DNS zone is not defined. Many of the specific implementation details are based on the implementation of Avahi [4], an open-source implementation that facilitates multicast DNS service discovery (mDNS/DNS-SD).

The privacy extension for mDNS/DNS-SD developed in [5], [6], [7] is one of the deciding factors for which these models have been developed. Most larger networks, such as campus networks, deactivate multicast by default, making it hard to scale experiments to larger scopes. Using a simulation model, this limitation can be eliminated to some extent, leaving the experiment limited only by available resources.

Our models allow researchers to design and evaluate both projects more rapidly using DNS, mDNS and extensions thereof. In this paper we provide an overview of the models' architectures and components to help researchers utilizing them. All models proposed in this paper are available as part of an open-source project hosted on GitHub[1].

## II. DNS MODEL

We implemented multiple components, each considering different aspects of the DNS architecture, including DNS servers, clients, and the functionalities needed to resolve and send queries. Figure 1 provides an overview of the interaction between the modules used within the DNS model. For comprehensibility the diagram does not include the relationships with lower level functions and structures needed for the implementation.

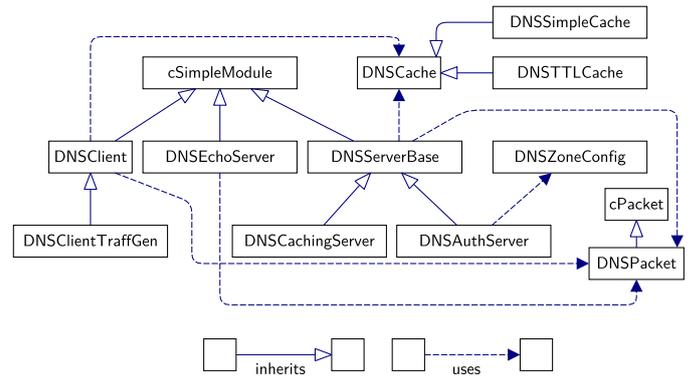

Fig. 1. Interaction diagram of the simple and compound modules implemented in the DNS model.

As shown, the model consists of the following components:

**DNSServerBase:** This class provides the basic capabilities a DNS server should support. For instance, iterative querying is done within the server base, as well as caching of records. To achieve this, the server can use the generic caching interface provided by `DNSCache` and can utilize different caching strategies depending on the implementation.

**DNSAuthServer:** This module extends the basic module `DNSServerBase` and adds the functionality of an authoritative DNS server, meaning `DNSAuthServer` is authoritative for a specific zone and answers accordingly. The initialization of a zone configuration is done using the `DNSZoneConfig` class. Not all configurations that are

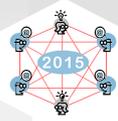

```
────────── Example Configuration ──────────
$TTL 86400 ; 24 hours, $TTL used for all RRs
ORIGIN uni-konstanz.de.
@ IN SOA pan.rz.uni-konstanz.de.
         hostmaster.uni-konstanz.de. (
           2003080800 ; sn = serial number
           172800     ; ref = refresh = 2d
           900        ; ret = update retry = 15m
           1209600    ; ex = expiry = 2w
           3600       ; nx = nxdomain ttl = 1h
           )
   IN  NS pan.rz.uni-konstanz.de.    ; in the domain
   IN  NS uranos.rz.uni-konstanz.de. ; slave
   IN  MX imap.uni-konstanz.de.      ; external mail
   IN  A  134.34.240.80              ; ip of origin
; server host definitions
pan.rz     IN A    134.34.3.3     ; this server
uranos.rz  IN A    134.34.3.2     ; the slave server
imap       IN A    134.34.240.42 ; mail server imap
www        IN CNAME proxy-neu.rz  ; test on
proxy-neu.rz IN A  134.34.240.80 ;
```

Fig. 2. Example zone configuration based on BIND[2] syntax.

typically possible with real world zone configurations, are currently implemented. Figure 2 shows a (working) sample configuration, based on the syntax used in BIND[2].

**DNSCachingServer:** This implementation can answer recursive queries by asking iteratively and performing cache lookups.

**DNSEchoServer:** The echo server is implemented according to [2] and provides the *echo domain* `.00.` and the *CCA* method `.cca..`

**DNSClient:** Using this module, DNS servers can be queried. For this purpose, the module provides a function `resolve`, which takes multiple arguments needed for the query, as well as callback handles that are called when the query has been performed. This enables modules, which using this implementation to perform some operation on the received data.

**DNSClientTraffGen:** For simple simulation purposes, this module can be used to perform basic queries. Therefore, it needs to read the desired queries from a file, the location of which can be specified as a parameter within the NED description of this module. It then randomly chooses from the queries and periodically sends them to the DNS servers.

**DNSCache:** This class represents an interface for caching DNS records. Currently we have two implementations, `DNSSimpleCache` and `DNSTTLCache`, the former evicting records from the cache randomly and the latter based on the records' lifetime.

**DNSPacket:** The `DNSPacket` extends the basic packet class `cPacket` and adds four lists of resource records, i.e. for the question, answer, authority and additional sections each. Furthermore, it adds the options of a DNS packet and the ID as proposed in [1].

This list is not comprehensive, since it does not include all parts that are crucial to this model's implementation. For a detailed description of all modules, classes and functions, see our official documentation[3].

[2] https://www.isc.org/downloads/bind/
[3] http://saenridanra.github.io/inet-dns-extension/doc/neddoc/index.html

*Capabilities:* At a high level the model's capabilities include the modeling of real world networks using DNS to resolve names, i.e. it supports the creation of hierarchical structures and is able to resolve names recursively and iteratively. To this end, a set of root servers must be defined, which are contacted by default if a resource's authoritative nameserver is unknown. Figure 3 shows an example DNS network and a client performing a query asking for `somehost.uni-konstanz.de`. Name compression is only considered when calculating the size of a packet, but not implemented for packets, since data structures are passed within the packet data structure and not as a payload consisting of only bytedata. However, following the example of the INET framework, we implemented a serializer for `DNSPackets` that currently serializes `A`, `AAAA`, `NS`, `PTR`, `SRV`, `CNAME`, and `TXT` records properly and performs name compression where it is permitted.

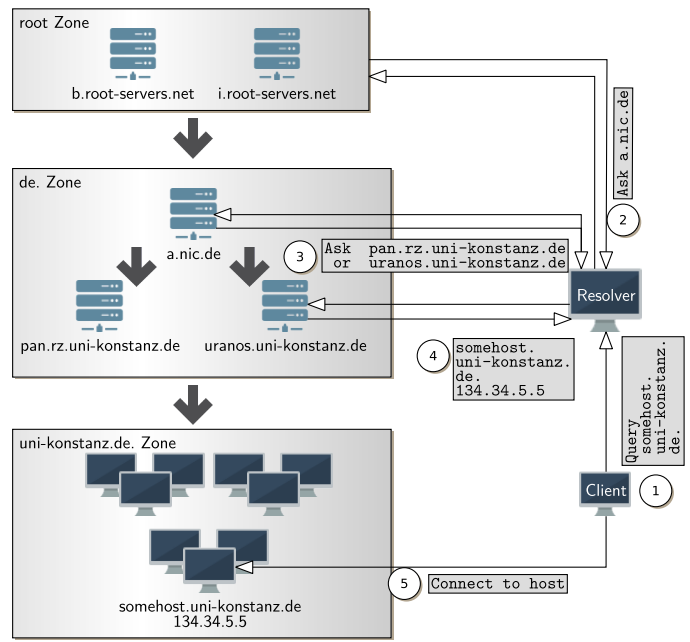

Fig. 3. This Figure shows a subset of an example network provided in the project source code, as well as the information flow in a live example.

*Implementation challenges:* OMNeT++ is well-suited for simulating the DNS protocol. The main difficulty resides in configuring a representative DNS network, and even more importantly, dynamically generating such networks. Another difficulty was analyzing and mapping existing rules of resolving and caching queries, since many of those rules are implementation-specific and not defined in the RFC. Providing extensibility and access in a generic way, as well as integrating the model within the INET framework to ease usage for researchers, were important concerns as well.

*Limitations:* Some functionalities have not been implemented, since they are not directly needed for the evaluations the models have been developed for:

- A DNS network has to be modeled manually. This includes defining the zone configurations for DNS servers

and providing IP addresses accordingly. It would be preferable to assign IP addresses for a known host within a zone after automatic configuration using the `IPv4NetworkConfigurator`.

- Bailiwick rules as described in [8] are currently not implemented.
- Currently the DNS servers only reply properly to `A`, `AAAA`, `NS`, `MX`, `CNAME` and `ANY` queries. Other operations can be easily implemented, as placeholders are provided at corresponding positions.
- There is no support for dynamic zone updates of DNS servers.
- Extensions such as DNSSec [9] are currently not implemented.

## III. MDNS MODEL

mDNS [3] provides local name resolution functionality. It is widely used in combination with DNS-SD [10] to provide zero configuration service discovery in local networks. The model proposed in this paper supports both mDNS and DNS-SD, one of the major goals being to evaluate the performance of mDNS/DNS-SD in networks in combination with the privacy extension and *Stateless DNS* [2]. Figure 4 provides an overview of how the different components in the model interact with each other.

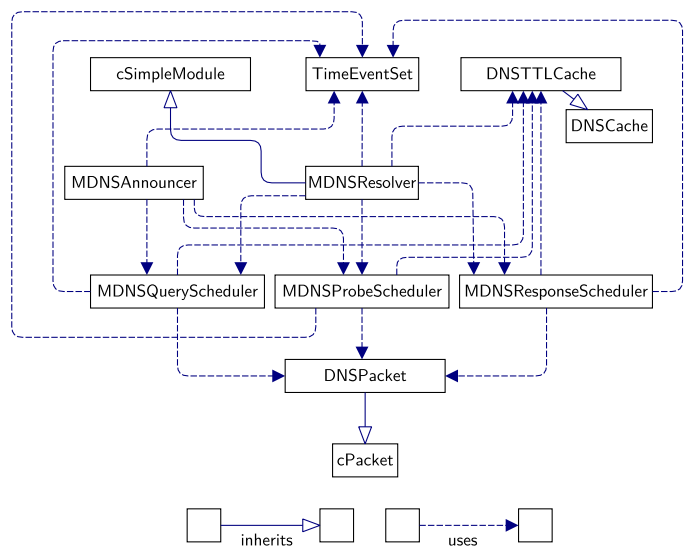

Fig. 4. Interaction diagram of the components implemented and used in the mDNS model.

As in Figure 1, only the most important components and their relations are shown, which are:

**MDNSResolver:** A simple module that essentially uses schedulers to query, probe and respond. It schedules *self-messages* according to the next event due in the set of time events and performs callbacks on elapsed events. It also handles the initialization of configured services, pairing data and private services.

**MDNSAnnouncer:** This class announces configured services to the network according to [3]. Hence, in a first step the services are probed for existence. If no conflict occurs, the service is announced to the network by sending unsolicited responses as defined in [3].

**MDNSProbeScheduler:** The probe scheduler probes the network with services that are to be announced, so that conflicts can be avoided. When a probe is posted and is not marked for immediate transmission, it is first put into a list and after a maximum of 250 ms, it is sent out along with other probes that need to be sent. Therefore, less packets are sent into the network. This timeframe also gives other devices the opportunity to respond on conflicting probes, that can be taken out of the schedule.

**MDNSQueryScheduler:** The query scheduler maintains queries that are to be sent out and additionally performs *duplicate question suppression* as defined in [3]. Additionally, *known answer suppression* is performed to further reduce the amount of traffic that is sent over the network.

**MDNSResponseScheduler:** In addition to maintaining response schedules, this class also performs *duplicate answer suppression* as defined in [3].

**TimeEventSet:** This class wraps a standard library container, more specifically an ordered set. The order is given by a `TimeEventComparator` that compares elements based on their expiry time. Since it is ordered, the head element of the set is always the next event due and since it is a set, events can be easily deleted and inserted, which would be more difficult with a priority queue. The time event set is used to maintain schedules and only set a single *self-message* based on the next due event, thereby improving the efficiency of the simulation. A time event is linked to a callback, so that a specific operation, such as checking if a probe can be sent out, can be performed when the event is due.

A component not shown in Figure 4, which is nevertheless crucial for evaluation, is the `MDNSNetworkConfigurator` module; it allows performing larger experiments without the need to configure resolvers manually. A network including this module is the *dynamic_mdns_network* example which is provided in the project source code. It can be configured to use a dynamic number of hosts, the number of services they use and different privacy related parameters that will be discussed in the context of the privacy extension.

*Capabilities:* At this time the model includes most functionalities described in [3]. In addition, it facilitates performing large experiments using dynamic parametrization. The startup procedure and announcement of static services is fully implemented.

*Implementation challenges:* Due to the restrictions set by [3] on how and when to send queries, probes and responses, scheduling, as it is done within the *Avahi Daemon*, is preferable. A simple solution would be to periodically set a *self-message* and check if an event is due. To improve efficiency, only one *self-message* is used based on the earliest event due and rescheduled in case of a new event that is due earlier. Another challenge is how to dynamically create and parameterize such networks, since it is hard to determine how many devices in a network use mDNS and DNS-SD and how many services are used. By varying parameters like the number of mDNS/DNS-SD hosts and the number of services they can use different scenarios can be evaluated.

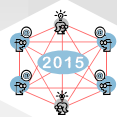

*Limitations:* There are some limitations and some functionalities currently not implemented:

- Shared resource records are not handled in separate data structures, as it is done in Avahi.
- Dynamic traffic generation is not implemented, meaning that resolvers will announce static services and periodically send unsolicited responses to reannounce the services, but will not query for services dynamically.
- Placeholders for internal module messages, initiating resolvers to query for services, are in place, but not yet implemented. These could be used by a compound module including a traffic generator to query for services and therefore simulate a more realistic network.
- Similar to the DNS model, not all operations defined by DNS and later extensions are supported for the mDNS model, for instance a query using the `AXFR` type.

## IV. PRIVACY EXTENSION

The idea of the privacy extension is to reduce the amount of information published with mDNS-SD and to reduce the amount of traffic transmitted by mDNS-SD. We only provide a brief overview of the privacy extension, since a detailed description can be found in [5], [6], [7].

To prevent private information being sent via multicast, the information is sent via a separate socket directly to trusted devices. The network parameters of this socket are offered and requested using a special meta-service that can distribute the desired information using alternative methods, e.g. our *Stateless DNS* technique [2]. Stateless DNS enables the use of mDNS-SD in networks where multicast is disabled because all information can directly be sent using unicast. While adding these features, the privacy extension is still fully backwards compatible. The presented models have been developed in order to measure the performance of mDNS-SD with and without the privacy extension, as well as using *Stateless DNS*. Using the `MDNSNetworkConfigurator` parameters can be varied to evaluate different scenarios. Parameters that can currently be modified include: number of resolvers, number of private resolvers, minimum and maximum number of friends and services. This already enables the simulation of large networks with mDNS-SD with a large number of resolvers, which could be used as an example of a network in an airport in which mDNS-SD is enabled.

Preliminary simulations have shown that our privacy extension significantly reduces multicast traffic and reduces network load by more than 50%. A sample of the results is shown in Figure 5.

## V. ONGOING AND FUTURE WORK

As part of our ongoing work, we want to add different traffic sources to the networks and evaluate how multicast in general and more specifically mDNS-SD affects the network performance of wireless networks. Additionally other concepts of reducing traffic generated by mDNS-SD are to be evaluated.

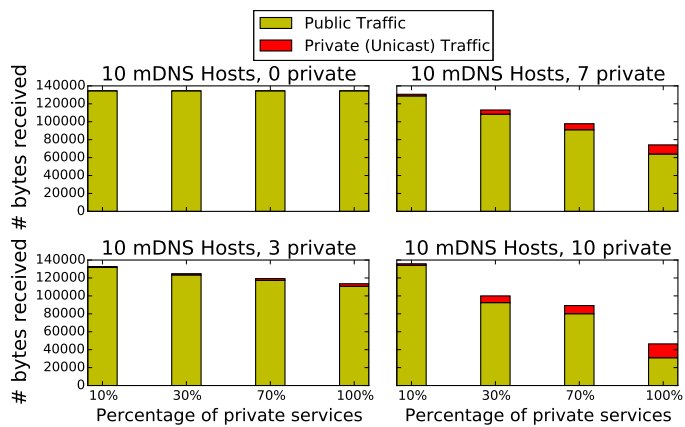

Fig. 5. Each subfigure shows a setup with ten mDNS hosts. The number of private hosts differs for each subfigure. The y-axis shows the number of bytes received. The x-axis shows the ratio of private services. With increasing privacy the traffic is reduced significantly.

Future work using the models presented in this paper may include:

- Dynamic generation of DNS networks.
- An implementation and evaluation of DNSSec [9].
- The implementation of host update protocols for DNS.
- Implementation and analysis of DNS caching (e.g. Bailiwick [8] rules) behavior to promote better caching rules and fine tune them towards performance and security.
- The evaluation of other experimental protocol extensions to DNS/mDNS without the need to test in the real world.
- Better integration of the models into the INET framework for ease of use.